\documentstyle[prl,aps,floats]{revtex}
\input	epsf
\font\latin=cmsy10
\def\guz{\hbox{\latin P$_{\rm G}$}}

\def\be{\begin{equation}}
\def\ee{\end{equation}}
\def\bea{\begin{eqnarray}}
\def\eea{\end{eqnarray}}

\def\ie{{\sl i.e.\ }}
\def\ea{{\sl et	al., }}
\def\ean{{\sl et al.\ }}
\def\tj{$t$--$J$\ }
\def\ibid{{\sl ibid.\ }}

\begin{document}
\draft
\twocolumn[\hsize\textwidth\columnwidth\hsize\csname@twocolumnfalse%
\endcsname

\title{
Fictitious flux	confinement:
magnetic pairing in coupled spin chains	or planes}

\author{Martin Greiter}
\address{Department of Physics,	Stanford University, Stanford, CA 94305,
greiter@quantum.stanford.edu}

\date{SU-ITP 98/20, cond-mat/9805112, May 10, 1998}

\maketitle

\begin{abstract}
The spinon and holon excitations of two-leg Heisenberg or lightly doped
$t$--$J$ ladders are shown to be bound in pairs	by string 
confinement
forces given approximately by the antiferromagnetic exchange energy
across the rungs, 
$F= J_\perp \langle {\bf S}_i {\bf S}_j	\rangle_\perp /b$.
These forces originate from the	fictitious flux	tubes associated with
the half-fermi statistics of the excitations.
It is conjectured that similar confinement forces,
determined by the antiferromagnetic exchange energy across the layers,
are responsible	for the	spin
gap and	the pairing of charge carriers in CuO superconductors.
\end{abstract}

]


1.\
Two-leg	Heisenberg or lightly doped $t$--$J$ ladders \cite{science}
can, according to a recent proposal \cite{lad},	be described approximately
by 
an RVB type spin liquid.  To be	precise, one assumes a magnetic
tight-binding model on the ladder, with	flux $\pi$ per plaquet
and hopping magnitudes $\tilde t$ along	the chains
and $\tilde t_\perp$ across the	rungs.	One then fills the lower band twice,
once with up-spin and once with	down-spin electrons.  Upon elimination
of doubly occupied sites via Gutzwiller	projection, one	obtains	a
reasonable approximation \cite{lad} to the ground state	of the Heisenberg
ladder with $J_\perp/J=\tilde t_\perp/\tilde t$.

Spinon and holon excitations for this liquid may be created either
via Anderson's projection technique \cite{phil}	or via midgap
states \cite{rokh}.  As	the topology of	the ladder dictates that midgap
states can only	be created in pairs, the second	method
automatically yields spinon-spinon bound states	(magnons) rather then
isolated spinons,
which reflects the fact	that the spinons or holons 
are confined in	pairs.

In the first part of this letter, I will explain the form and	origin of
the confinement	forces between the spinons and holons of the ladder,
and calculate the spinon mass and the bound state resonances
based on a heuristic identification of the spin	gap of the
Heisenberg ladder as the zero point energy of the string oscillator.
In the second part,
I will postulate similar confinement forces in systems of (weakly) coupled
magnetic planes, make some assumptions regarding both the nature of
the spin liquid	in the planes and the confinement forces
due to the interplane coupling,	and obtain an estimate for the spin
and charge gaps	in bilayer CuO superconductors 
(25 meV	and 28 meV, respectively, for YBa$_2$Cu$_3$O$_{6+x}$).
I conclude with	several	comments on high-$T_c$ superconductivity.

\begin{figure}
\centerline{\epsfbox{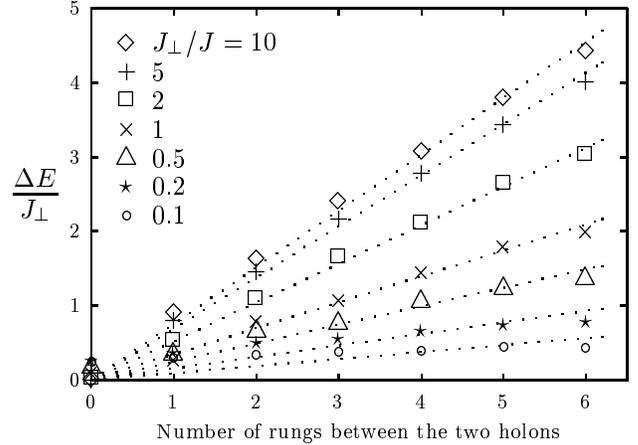}} \vspace{8pt}
\caption{The confinement energy, defined here as the
energy of a pair of holons measured relative to the
exact ground state energy of a pair of holes, as a function
of the number of rungs between the holons or holes for a $2\times 8$
ladder with open boundary conditions and
$\tilde t_\perp/\tilde t=0.75 (J_\perp/J)^{0.5}$ 
for $J_\perp/J\leq 1$
and $\tilde t_\perp/\tilde t=(J_\perp/J)^{0.6}$	for $J_\perp/J\geq 2$,
\ie such that the ratios 
$\langle \vec S_i \vec S_j \rangle_\perp /
\langle \vec S_i \vec S_j \rangle_\parallel $ 
of the trial wave
functions match	those of the exact ground states.
One holon was localized at the end of
the ladder and the other localized at various positions	on the
same chain, as illustrated in fig.\ 3a; 
a finite size
correction has been taken into account for $J_\perp/J=0.2$ and $0.1$.
} 
\label{f:one}
\end{figure}

\begin{figure}
\centerline{\epsfbox{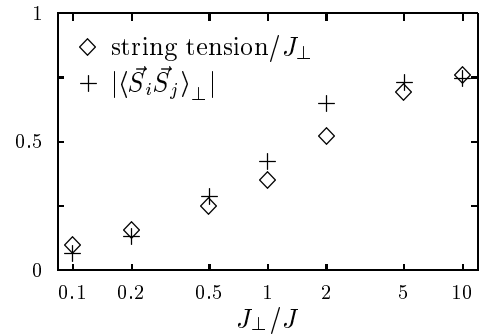}}\vspace{5pt}
\caption{The string tension in units of	$J_\perp$
given by the slope of the dotted lines fitted
though the data	points in fig.\	1 
in comparison with the
spin correlations $|\langle \vec S_i \vec S_j \rangle_\perp |$
across the rungs in the	ground state
as taken from table
I of [2] 
for different ratios $J_\perp/J$.
The discrepancy	around $J_\perp/J=2$ arises from the
enhanced correlations $|\langle	\vec S_i \vec S_j \rangle_\parallel |$
along the chains in the	presence of the	holons,	which have been
neglected in (2). 
}
\label{f:five}
\end{figure}

\begin{figure}
\centerline{\epsfbox{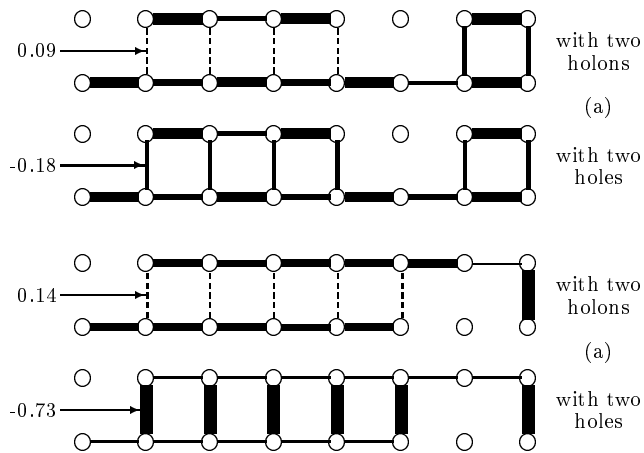}}\vspace{12pt}
\caption{Spin correlations $\langle \vec S_i \vec S_j \rangle$
between	nearest	neighbors in the presence of two holons	compared with
two holes for
two representative configurations of
a $2\times 8$ ladder with open boundary	conditions and
(a) $J_\perp/J=0.5$ or (b) $J_\perp/J=5$.
The conventions	are adopted from White and Scalapino [6] 
the thickness of the lines is proportional to
$|\langle \vec S_i \vec	S_j \rangle |$;	solid lines indicate
antiferromagnetic, dotted lines	ferromagnetic correlations.
The state containing two holons
was obtained from Gutzwiller projected magnetic	bands using
Anderson's method for constructing spinons; the state
containing two holes is	just the exact ground state of the Heisenberg
ladder with two	static vacancies at the	positions indicated.
}
\label{f:four}
\end{figure}

2.\
In order to determine the functional form and strength of
the confining potential	between	the
spinon or holon	excitations of the two-leg Heisenberg ladder,
we create two holons at	sites $i$ and $j$
via Anderson's projection technique \cite{phil},
\be
|\psi_{i,j} \rangle = c_{i\uparrow}c_{j\downarrow}\,
\guz\, c^\dagger_{i\uparrow}c_{j\uparrow}|\psi_{\rm SD}\rangle,
\ee
where $|\psi_{\rm SD}\rangle$ is the Slater determinant	ground state
obtained by filling the	lower magnetic tight-binding band twice,
and numerically	compare	the energy expectation value of	this configuration
to the energy of the exact ground state	for a Heisenberg ladder	with
two stationary holes at	these sites \cite{strictly}.  The results
are shown fig.\	\ref{f:one} and	\ref{f:five}; we find a	linear potential
\be
V(x)=F |x| \hbox{\hspace{12mm}}	
F\approx J_\perp |\langle {\bf S}_i {\bf S}_j \rangle_\perp |/b
\label{e:v}
\ee
where $x$ is the distance between the spinons or holons
in the direction of the	chains,
$\langle {\bf S}_i {\bf	S}_j \rangle_\perp$ the	spin correlation
across the rungs, and $b$ the bondlength along the chains.
The confinement	energy is hence	approximately equal to the
antiferromagnetic exchange energy across the rungs between the spinons;
the reason for this emerges from a comparison of the individual
spin correlations on each link,	as reproduced in fig.\ \ref{f:four}
for two	typical	configurations:	 an invisible string between the
holons destroys	the antiferromagnetic correlations on all the rungs
between	them.

\begin{figure}
\centerline{\epsfbox{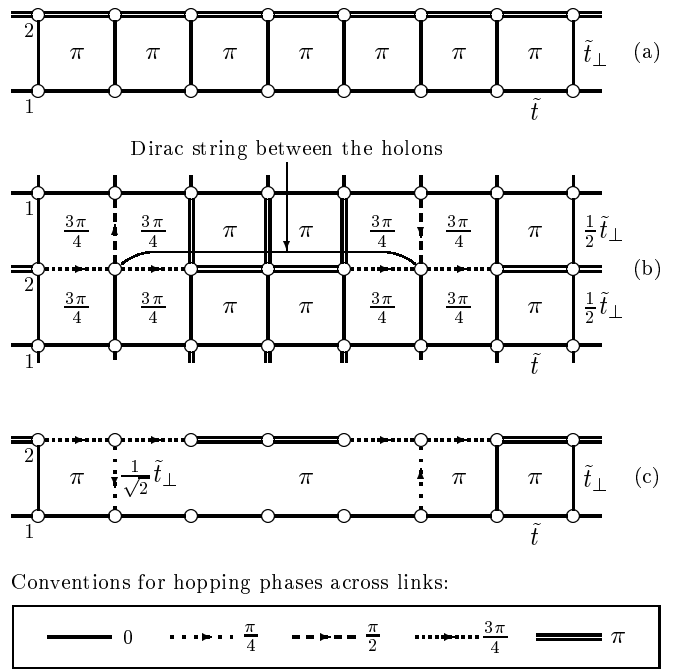}}\vspace{10pt}
\caption{
Magnetic tight-binding configurations before Gutz\-willer projection for
(a) the	ground state of	the \tj	ladder at half-filling
and (b,c) in the presence of two holons	at a distance of 4 lattice
spacings.  The holons in (b) were created following the	procedure
for the	Kalmeyer-Laughlin chiral spin liquid [7] 
suggested in [2] 
after mapping the	flux ladder into a flux	lattice	subject	to
a periodic boundary condition with a periodicity of only two
lattice	spacings in $y$-direction.  The	lattice	is subsequently
reconverted into a ladder (c).	Note that the Dirac string
annihilates all	the hopping terms across the rungs between the
holons,	while the hopping magnitudes along the chains remain unaffected.
}
\label{f:thirteen}
\end{figure}

The origin of this string is explained in fig.\	\ref{f:thirteen}:
the fictitious flux tube associated with the half-fermi	statistics
\cite{wilc} of the spinons or holons \cite{chiral,hald},
which manifests	itself in an adjustment	by $\pi$ \cite{rokh,lad}
of the fictitious flux through the adjacent plaquets in	the
magnetic tight-binding model before Gutzwiller projection, 
effectively annihilates	the hopping terms on the rungs between them.

%
%

3.\
If we were to know the effective mass of the spinon, we	could calculate
the resonances of the string oscillator
\be
(-{1\over2m_{\rm red}}\nabla^2 + F |{\bf r}|\,)\,\psi_n({\bf r})=
E_n\psi_n({\bf r}),
\label{e:dfgl}
\ee
where $m_{\rm red}={1\over2}m_{\rm sp}$	is the reduced mass of a
spinon pair, and compare the ground state energy
to the spin gap	in the Heisenberg
ladder,	which is just the energy required to create a spinon-spinon
bound state; apart from	a possible correction due to a chemical	potential
for spinons which we neglect, these should be equal.  In the present case,
we use the know	value $\Delta\approx J_\perp/2$
for the	spin gap \cite{ladgap} to calculate the	spinon mass
and the	spinon-spinon resonances in the	weak coupling regime $J_\perp<J$.
In one dimension, the solutions	of (\ref{e:dfgl})
are given in terms of the Airy function	${\rm Ai}(x)$ \cite{airy},
\be
\psi_n(x)=\left({x\over|x|}\right)^n
{\rm Ai}\left({|x|\over	x_0}-\lambda_n\right), \hbox{\hspace{2mm}}
E_n=F x_0 \lambda_n,
\label{e:airy}
\ee
where $x_0=1/(2m_{\rm red} F)^{1/3}$ is	the characteristic lengthscale
of the oscillator, and $\lambda_n$ are the extrema or zeros of
${\rm Ai}(-x)$ for $n$ even or $n$ odd,	respectively, which are	listed in
table \ref{t:eleven}.  Equating	$E_0=\Delta$, 
we obtain
\be
m_{\rm sp}={F^2\lambda_0^3\over	E_0^3}=
{8\,|{\langle {\bf S}_i	{\bf S}_j \rangle}_\perp |^2\, \lambda_0^3
\over J_\perp b^2} = 3.25 {J_\perp\over	J^2 b^2},
\label{e:msp}
\ee
where we have approximated
$|\langle {\bf S}_i {\bf S}_j \rangle_\perp |\approx 0.62 J_\perp/J$
according to table
I of \cite{lad}
for weakly coupled ladders.
The spacing of the eigenvalues $\lambda_n$ implies
that the internal resonance frequencies	of the
spinon-spinon bound states or magnons are comparable but slightly
higher than the	energies required to create a second or	third magnon;
this proximity and
the fact that they presumably decay rapidly into several magnons
renders	a potential observation	of the resonances difficult if not impossible.

It is now easy to estimate the size of the magnon.  Using
${\rm Ai}(|x|-\lambda_0)\approx	{\rm Ai}(-\lambda_0)\exp (-{1\over 3}x^2)$,
we write the ground state
\be
\psi_0(x)=
\exp (-{x^2\over 2\xi^2})
\hbox{\hspace{1mm} with\hspace{2mm}}
\xi={\textstyle	\sqrt{3\over 2}} x_0 
= 0.97 {J\over J_\perp}b.
\label{e:xi}
\ee
This result illustrates
why the	spin gap in the	weak coupling regime can be 
${1\over2}J_\perp$ while the
antiferromagnetic exchange energy across each rung is only of order
${1\over2}J_\perp^2/J$:	the number of decorrelated rungs
is of order $J/J_\perp$.

This calculation can also be applied to	spinon-holon bound states
(holes)	in the ladder;
the only difference is that the	reduced	mass in	(\ref{e:dfgl}) and
(\ref{e:airy}) is replaced by
\be
{1\over	m_{\rm red}}={1\over m_{\rm sp}}+{1\over m_{\rm	h}},
\hbox{\hspace{2mm} where\hspace{3mm}}
m_{\rm h}={1\over 2t_{\rm eff}b^2}
\label{e:mh}
\ee
is the effective mass of the holon; according to table II of \cite{lad},
$t_{\rm	eff}={1\over 2}	E_{t_\parallel}	=0.77\,	t$
for $J_\perp=J$	and $t_{\rm eff}= 0.95\, t$ in the weak	coupling limit
$J_\perp \ll J$.  The values for the  energies $E_n$
and the	size $\xi$ of the bound	state are those	given in (\ref{e:airy})
and (\ref{e:xi}) above multiplied by
\be
\mu=
\left({1\over2}+{m_{\rm	sp}\over 2m_{\rm h}}\right)^{1/3}
=\left({1\over2}+ 3.25 {t_{\rm eff} J_\perp\over J^2}\right)^{1/3}
\label{e:mu}
\ee

\begin{table}
\begin{tabular}{c|cc|ccc}
\hspace{6mm} & \multicolumn{2}{c|}{\hbox{1d string\hspace{4mm}}}   
     & \multicolumn{3}{c}{2d string oscillator} \\
$n$  &\hspace{0mm}$\lambda_{2n}$  \hspace{0mm} 
     &\hspace{0mm}$\lambda_{2n+1}$\hspace{4mm} 
     &$\lambda_{n0}$ &$\lambda_{n1}$ &$\lambda_{n2}$ \\\hline
0  & 1.0188  & 2.3381\hspace{4mm}  & 1.7372  & 2.8721  & 3.8175 \\
1  & 3.2482  & 4.0879\hspace{4mm}  & 3.6702  & 4.4930  & 5.2629 \\
2  & 4.8201  & 5.5206\hspace{4mm}  & 5.1697  & 5.8671  & 6.5415 
\end{tabular}
\vspace{5pt}
\caption{The lowest (dimensionless) energy eigenvalues for the linear
oscillator in one and two dimensions.}
\label{t:eleven}
\end{table}

4.\
I will now turn	to the more speculative	part of	this letter, and explain
part of	my thinking on CuO superconductivity.
To begin with, I make the following assumptions:
\hspace{1mm} 1.\ The ground state of the
two-dimensional	\tj model at the relevant hole dopings is a spin
liquid,	which supports spinons and holons as elementary	excitations
\cite{phil}.
(My understanding \cite{martin}	of this	liquid is that it is
a liquid in both the spin degrees of freedom and in the	non-relativistic
plaquet	chiralities \cite{wwz};	this {\it chirality liquid} may	be
seen as	a significant generalization of	Laughlin's chiral spin liquid
\cite{chiral}, which is	a liquid in the	spins but effectively aligns
the chiralities	and thereby violates the discrete symmetries P and T.
These symmetries are preserved in my construction.  The	spinon and
holon excitations supported by the chirality liquid carry a chirality
quantum	number,	which can be $+$ or $-$; this number determines	the
sign of	the winding phases associated with their half-fermi statistics.)
\hspace{1mm} 2.\
The spinon and holon masses in a system	of coupled \tj planes are comparable
to their values	in a system of coupled chains estimated	above.
\hspace{1mm} 3.\
The analysis above, including all the implicit assumptions made, is correct.

The main difference between fictitious flux confinement	in a system
of coupled planes as compared to coupled chains	is that	the
one-dimensional	array of
decorrelated rungs is replaced by a puddle of decorrelated
interplane links.
To see this, imagine several ladders
(which are not necessarily straight) embedded into a system of coupled
planes such that the rungs align with interplane links,	and connect
two spinon sites $i$ and $j$ along various paths in the	planes
with these ladders; the	fictitious flux	connecting the spinons
will then destroy the correlations across all the rungs	on each	ladder.

The simplest estimate for the strength of the confining	force is
to assume a circular droplet of	decorrelated links
between	the spinons, with a diameter given by the spinon-spinon
distance $r$.
(This is presumably not	a valid	approximation for large	spinon separations,
but may	be reasonable for the ground state of the oscillator.)	It
yields a harmonic potential
\be
V(r)={\pi r^2\over 4b^2}
J_\perp	|{\langle {\bf S}_i {\bf S}_j \rangle}_\perp |
\equiv {\textstyle {1\over2}}Dr^2.
\ee
The ground state energy	of the spinon-spinon bound state is hence given	by
\be
E_0=\sqrt{D\over m_{\rm	red}}
=\sqrt{{\pi\over 3.25} |\langle	{\bf S}_i {\bf S}_j \rangle_\perp | J^2}
=0.77 \sqrt{J_\perp J},
\ee
where we have used (\ref{e:msp}) and
$|\langle {\bf S}_i {\bf S}_j \rangle_\perp |\approx 0.62 J_\perp/J$.
This estimate for the spin gap \cite{ong} in YBa$_2$Cu$_3$O$_{6+x}$
is directly related to the optical magnon gap in the ordered
antiferromagnet	YBa$_2$Cu$_3$O$_{6.2}$,	which has been measured by
inelastic neutron scattering \cite{neut}:
\be
E_{\rm opt.}=2\sqrt{J_\perp J}\approx 70\ \hbox{meV},
\ee
which yields a spin gap	of 27 meV.
The charge gap is just the gap to create a spinon-holon	bound state:
substituting $J=120\ \hbox{meV}$, $J_\perp=10\ \hbox{meV}$ 
$t=500\	\hbox{meV}$, and $t_{\rm eff}=0.9\,t$ for YBa$_2$Cu$_3$O$_{6+x}$
into (\ref{e:mu}), we obtain
31 meV.

\begin{figure}
\centerline{\epsfbox{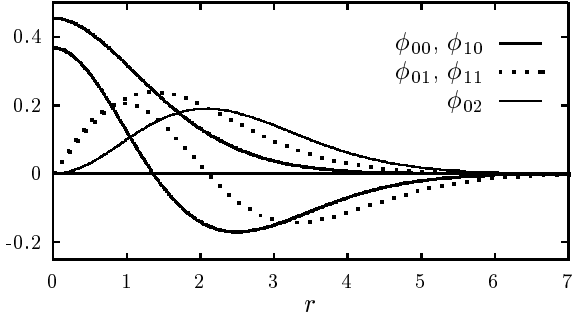}}\vspace{5pt}
\caption{Radial	wave functions for the lowest energy eigenstates of the
linear oscillator in two dimensions.}
\label{f:fivteen}
\end{figure}

The microscopic	details	of the
chirality liquid \cite{martin} yield a linear potential
as an estimate for the confining force,
\be
F_{\rm 2d}={1\over b}{\sqrt{E_{00}
J_\perp	|\langle {\bf S}_i {\bf	S}_j \rangle_\perp | \over 2}},
\label{e:fchi}
\ee
where $E_{00}$ is the spin gap given by	the ground state energy	of the
string oscillator (\ref{e:dfgl}) with string tension
$F_{\rm	2d}$ in	two dimensions.	 This oscillator has to	be solved
numerically; the solutions are
\be
\psi_{n\ell}(r,\varphi)=
e^{\pm i \ell\varphi}\phi_{n \ell}
\left({r\over r_0}\right),  \hbox{\hspace{2mm}}
E_{n \ell}=F_{\rm 2d} r_0 \lambda_{n\ell},
\label{e:psichi}
\ee
with $r_0=1/(2m_{\rm red} F_{\rm 2d})^{1/3}$, $\lambda_{n\ell}$
as listed in table \ref{t:eleven}, and $\phi_{n	\ell}(r)$ as
plotted	in fig.\ \ref{f:fivteen}.

Solving	(\ref{e:fchi}) and (\ref{e:psichi}) for	$E_{00}$ yields
\be
E_{00}=\sqrt{J_{\perp} |\langle	{\bf S}_i {\bf S}_j \rangle_\perp |
\lambda_{00}^3 \over 4m_{\rm red}}
={\textstyle \sqrt{{1\over 2}J_\perp J}},
\ee
where we have used (\ref{e:msp}) and
$|\langle {\bf S}_i {\bf S}_j \rangle_\perp |\approx 0.62 J_\perp/J$
once more.
This yields a spin and charge gap of 25	meV and	28 meV for
YBa$_2$Cu$_3$O$_{6+x}$.

It should of course be born in mind that these numbers are only
rough estimates; many important	details, including the $d$-wave
symmetry of the	superconducting	order parameter, have not been taken
into account here.

5.\
I wish to conclude with	a few general observations relating the	circle of
ideas drawn above to high-$T_{\rm c}$ superconductivity.

(a)
The fictitious flux confinement	forces are not
responsible for	the pairing of the charge carriers
in a conventional sense, as the	charge
superfluid forms an integral part of the spin liquid within the
planes;	in particular, the interplane correlations
are not	destroyed by the presence of charge carriers
in the superconducting state at	$T=0$.	The charge gap estimated
above, however,	corresponds roughly \cite{ilt} to the gap required
to create an ``unpaired'' hole,	which does not participate in the superfluid.
The confinement	forces are thus	responsible for	the robustness of
the pairing, including its persistence at finite temperatures, but not
for the	pairing	itself \cite{wieg}.

(b)
The strength of	the magnetic pairing force is determined by the
the magnetic energy stored in the structure between the	layers,	or
more precisely by the amount this energy is raised by the fictitious
flux, which is only in multilayer materials directly related to	the
coupling between the layers.  The mechanism hence allows for strong
pairing	forces in single layer materials; we expect those, however,
to be crucially	dependent on the magnetic properties of	the structure
between	the planes.

(c)
My thinking on the spin	gap phase coincides with that of Emery and Kivelson
\cite{emery}. 

(d)
The superconductivity in C$_{60}$ is, according	to my thinking,
due to Dirac quantization of the fictitious flux through the surface of
each molecule \cite{buck,thbuck,martin}.

I am deeply grateful to	S.\ Chakravarty, 
A.\ Kapitulnik,	S.\ Kivelson, 
but especially R.B.\ Laughlin for many illuminating discussions.
This work was supported	through	NSF grant No.~DMR-95-21888.
Additional support was provided	by the NSF MERSEC Program through
the Center for Materials Research at Stanford University.


\begin{references}
\bibitem{strictly}
Strictly	speaking, the exact ground state contains a small admixture
of a state with	two holons, which we have neglected here as it tends
to zero	as the coupling	$J_\perp/J$ gets stronger
or as $i$ and $j$ get separated	further.
\end{references}
\end{document}